\documentclass[aps,prb,twocolumn,superscriptaddress,longbibliography]{revtex4-1}
\usepackage{graphicx}
\usepackage{latexsym}
\usepackage{amssymb}
\usepackage{amsmath}
\usepackage{amsfonts}
\usepackage{upgreek}
\usepackage{bm}
\usepackage{bbold}
\usepackage{multirow}
\usepackage{color}
\usepackage[T1]{fontenc}
\usepackage{hyperref}
\hypersetup{
colorlinks = true,
linkcolor = [rgb]{0.70,0.13,0.13},
citecolor = [rgb]{0.13,0.55,0.13},
urlcolor  = [rgb]{0.25, 0.41, 0.88}}
\newcommand{\ket}[1]{|#1\rangle}
\newcommand{\bra}[1]{\langle#1|}

\begin{document}

\title{Dynamical quantum phase transitions in a spin chain with deconfined quantum critical points}
\author{Gaoyong Sun}
\thanks{Corresponding author: gysun@nuaa.edu.cn}
\affiliation{College of Science, Nanjing University of Aeronautics and Astronautics, Nanjing, 211106, China}
\author{Bo-Bo Wei}
\thanks{Corresponding author: weibobo@cuhk.edu.cn}
\affiliation{School of Science and Engineering, The Chinese University of Hong Kong, Shenzhen, Shenzhen 518172, China}
\affiliation{Center for Quantum Computing, Peng Cheng Laboratory, Shenzhen 518055, China}

\begin{abstract}
We analytically and numerically study the Loschmidt echo and the dynamical order parameters in a spin chain with a deconfined phase transition
between a dimerized state and a ferromagnetic phase. For quenches from a dimerized state to a ferromagnetic phase, 
we find that the model can exhibit a dynamical quantum phase transition characterized by an associating dimerized order parameter. 
In particular, when quenching the system from the Majumdar-Ghosh state to the ferromagnetic Ising state, we find an exact mapping into the classical Ising chain
for a quench from the paramagnetic phase to the classical Ising phase by analytically calculating the Loschmidt echo and the dynamical order parameters. 
By contrast, for quenches from a ferromagnetic state to a dimerized state, 
the system relaxes very fast so that the dynamical quantum transition may only exist in a short time scale. 
We reveal that the dynamical quantum phase transition can occur in systems with two broken symmetry phases and the quench dynamics may be independent on equilibrium phase transitions.
\end{abstract}

\maketitle

\section{Introduction}
Understanding the behaviors of many-body systems out of equilibrium is a central problem of research in physics \cite{eisert2015quantum}.
Dynamical quantum phase transitions (DQPTs) \cite{heyl2013dynamical,heyl2018review} are proposed to occur at critical times $t_c$ during the real time evolution with nonanalyticities of the rate function
after a sudden quench of the system across equilibrium quantum critical points. 
Contrary to conversional classical (quantum) phase transitions driven by temperature (magnetic field or pressure), 
DQPTs are considered as new types of phase transitions driven by time. There has been a lot of interest in the study of DQPTs, 
including critical properties \cite{heyl2015scaling,karrasch2017dynamical,chichinadze2017critical,khatun2019boundaries,wu2020nonequilibrium,wu2020dynamical,wu2019dynamical,trapin2020unconventional}, 
dynamical order parameters \cite{budich2016dynamical,sharma2016slow,bhattacharya2017emergent,heyl2017dynamical}, 
and spontaneously broken symmetries \cite{heyl2014dynamical,canovi2014first,fogarty2017dynamical,weidinger2017dynamical,huang2019dynamical,feldmeier2019emergent}, etc.
The realizations of DQPTs have been performed in a large number of experiments based on different platforms \cite{jurcevic2017direct,flaschner2018observation,xu2020measuring,wang2019simulating,tian2019observation,guo2019observation,nie2019experimental,tian2020observation}.

On the other hand, the deconfined quantum critical point (DQCP) was originally inroduced as 
a second-order quantum phase transition between the valence bond solid (VBS) state
and the antiferromagnetic (AF) Ne\'{e}l phase \cite{Senthil2004,Senthil2004PRB,Sandvik2007,Shao2016}.
The lattice symmetry is broken for the VBS phase, while the spin symmetry is broken for the AF phase \cite{Senthil2004}.
A transition between two different broken symmetry phases is usually considered as a first-order transition instead of second-order according to Landau-Ginzburg-Wilson theory \cite{Wilson1974,Wilson1975}.
Hence, in two-dimensional models, whether the phase transition is a DQCP or a weakly first-order is still under debate \cite{Shao2016}.
Contrary to two-dimensional systems, recent numerical results \cite{Jiang2019,Roberts2019,huang2019emergent,luo2019intrinsic,sun2019fidelity,mudry2019quantum} strongly support a continuous second-order transition 
with the conversional finite-size scaling \cite{luo2019intrinsic,sun2019fidelity} in one-dimensional models.
In this paper, we will focus our study on DQPTs in a one-dimensional spin chain with DQCPs.

The previous studies on DQPTs with spontaneously broken symmetry phases have been carried out in many systems
(i.e. discrete $\mathbb{Z}_2$ symmetries \cite{heyl2014dynamical,huang2019dynamical}, broken continuous symmetries \cite{weidinger2017dynamical}, lattice symmetry breaking \cite{feldmeier2019emergent},  etc).
Recently, the DQPTs were investigated in two-dimensional quantum dimer model with VBS phases by considering quenches across a Berezinskii-Kosterlitz-Thouless (BKT) transition 
and a first-order transition \cite{feldmeier2019emergent}.
However, to the best of our knowledge, whether DQPTs can occur in systems after a quench across a DQCP is so far less known.  
In the following, we will investigate DQPTs in systems with a quench between two broken symmetry phases based on the Loschmidt echo and dynamical order parameters.
We show that DQPTs can occur and be characterized by a dimerized order parameters for quenches from VBS phases to ferromagnetic (FM) phases.
More importantly, we find that DQPTs in spin chain with DQCPs for the quench from the Majumdar-Ghosh phase to the classical Ising phase can be mapped to DQPTs in the Ising chain
for the quench from the paramagnetic phase to the classical Ising phase.

This paper is organized as follows. 
In Sec.\ref{sec:LE}, we introduce the concept of Loschmidt echo. 
In Sec.\ref{sec:model}, we discuss the spin chain model with DQCPs we used. 
In Sec.\ref{sec:Isingchain}, we review the DQPTs in the transverse field Ising chain.
In Sec.\ref{sec:Results}, we present the main results of this paper. 
In Sec.\ref{sec:Con}, we summarize.

\begin{figure}
\includegraphics[width=8.8cm]{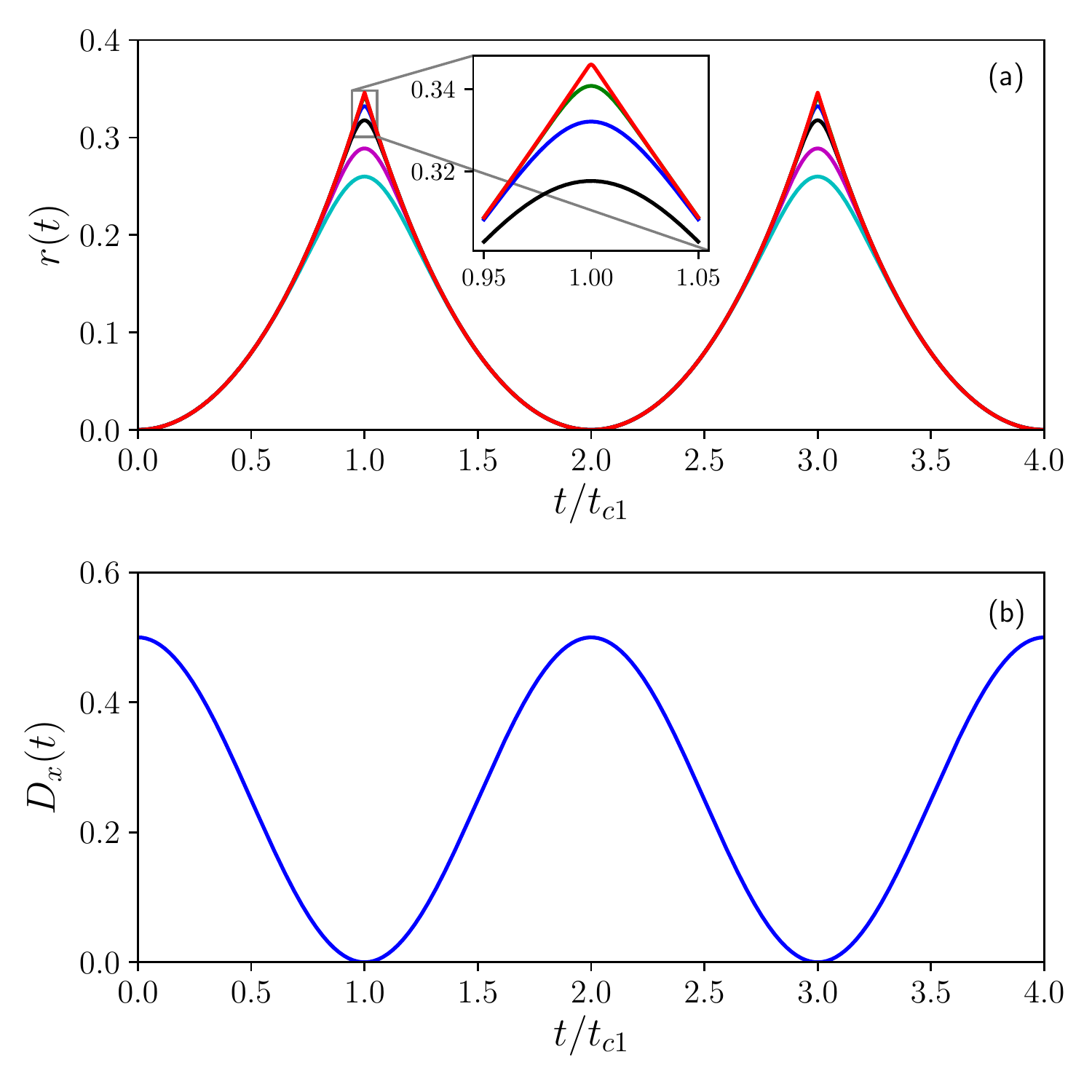}
\caption{(Color online)
Dynamics from the initial Majumdar-Ghosh state to the quenched classical Ising chain of Eq.(\ref{Ham}) with time $t/t_{c1}$ and $t_{c1}=\pi/4$. 
(a) Rate function $r(t)$ for $N=2400, 240, 96, 48, 24, 16$ lattice sites from top to bottom.  
The inset figure in (a) shows the finite-size effects near the first critical time for $L = 2400, 240, 96, 48$ from top to bottom.
(b) Dimerized order parameters $D_x(t)$ with the same parameters as (a). The results are the same as classical Ising model with $N^{\prime}=N/2$, 
where the dimerized order parameters $D_x(t)$ corresponds to the $\langle\sigma^x(t)\rangle$. 
}
\label{MGfig}
\end{figure}

\section{Loschmidt echo}
\label{sec:LE}
Given an initial quantum state $\ket{\psi_{0}}$, the Loschmidt amplitude $G(t)$ is defined as the overlap between the initial state $\ket{\psi_{0}}$ and its time evolved states $\ket{\psi(t)} = e^{-iHt} \ket{\psi_{0}}$,
\begin{align}
G(t) = \bra{\psi_0} e^{-iHt} \ket{\psi_0},
\label{LE:LA}
\end{align}
where $H$ is the quench Hamiltonian governing the time evolution of the system. The Loschmidt amplitude can be regarded as a dynamical counterpart to the partition function.
Thus we can define the return rate function,
\begin{align}
r(t) = -\frac{1}{N} \ln L(t),
\label{LE:RF}
\end{align}
as an analogy of the free energy of the classical systems \cite{heyl2013dynamical,heyl2018review}. 
Here $L(t)=|G(t)|^2$ is the Loschmidt echo,  $N$ is the system size. The rate function will exhibit nonanalytical behaviors (such as with a kink structure) at critical times.

\section{Model}
\label{sec:model}
We consider the following spin chain model for the nonequilibrium dynamics proposed recently in \cite{Jiang2019,Roberts2019,huang2019emergent,luo2019intrinsic,sun2019fidelity}
\begin{align}
H = \sum_{j} {}& (-J_{x} \sigma^{x}_{j} \sigma^{x}_{j+1} - J_{z} \sigma^{z}_{j} \sigma^{z}_{j+1} \nonumber \\
    {}& + K_{x} \sigma^{x}_{j} \sigma^{x}_{j+2}+K_{z} \sigma^{z}_{j} \sigma^{z}_{j+2}).
\label{Ham}
\end{align}
Where $\sigma^{x}_{j}, \sigma^{z}_{j}$ are the Pauli matrices at the $j$-th sites, $J_{x} \ge 0, J_{z} \ge 0$ and $K_{x} \ge 0, K_{z} \ge 0$ 
are the nearest neighboring and the next-nearest neighboring coupling constants, respectively.
The model has the $\mathbb{Z}_2 \times \mathbb{Z}_2$ symmetry, the translation symmetry and the inversion symmetry \cite{Jiang2019}.
The system undergoes a second-order quantum phase transition, known as DQCP, between the VBS dimerized phase and the FM phase.
We note that: (1) for $J_{x}=J_{z}=1$, $K_{x}=K_{z}=1/2$, the ground state is the Majumdar-Ghosh state \cite{Roberts2019}, 
(2) for $J_{x}=K_{x}=K_{z}=0$, $J_{z}=1$, the model is reduced to be the classical Ising model.
In the following, we will study the quench dynamics from the VBS phase to the FM phase and vice versa, respectively with periodic boundary conditions (PBC).

\begin{figure}
\includegraphics[width=8.8cm]{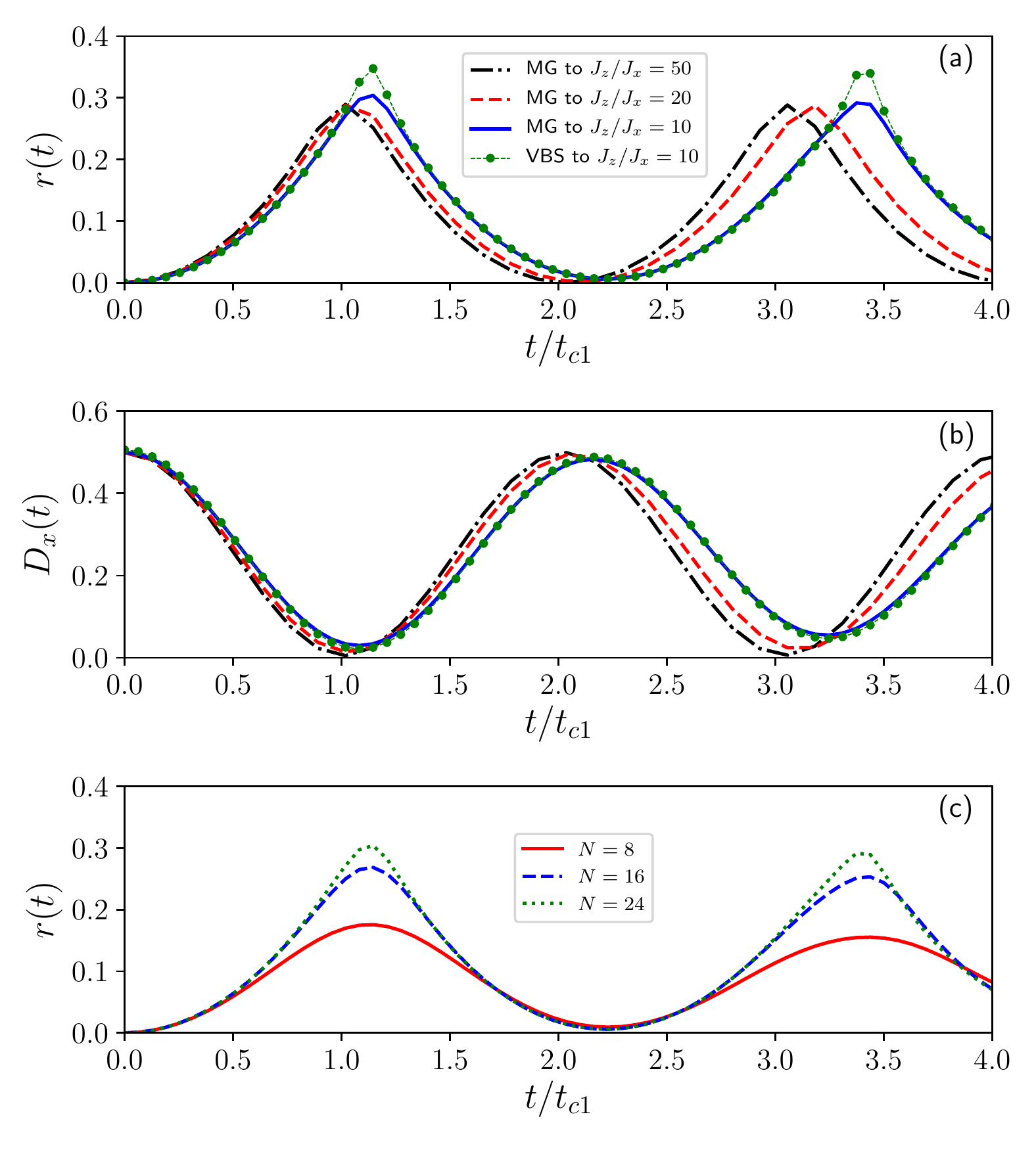}
\caption{(Color online)
Dynamics from the initial VBS phases to FM phases of Eq.(\ref{Ham}) with time $t/t_{c1}$ and $t_{c1}=\pi/4$. 
(a) Rate function $r(t)$ of $N=24$ lattice sites for the quench from Majumdar-Ghosh (MG) phase at $J_z/J_x=1$ to FM phase at $J_z/J_x=50$ (black dash-dot line), 
$J_z/J_x=20$ (red dashed line), $J_z/J_x=10$ (blue solid line), respectively.
The green filled-circle symbols denote the data for the quench from the VBS phase at $J_z/J_x=1.05$ to FM phase at $J_z/J_x=10$ with $N=24$ lattice sites. 
The results change little compared to that for the quench from MG state to FM phase at $J_z/J_x=10$.
(b) Dimerized order parameters $D_x(t)$ with the same parameters as (a). 
(c) The finite-size scaling of rate function $r(t)$ for the quench from MG phase to FM phase at $J_z/J_x=10$ with $N=8$ (red solid line), $N=16$ (blue dashed line), $N=24$ (green dotted line) lattice sites.
We rescale the Hamiltonian by choosing $J_z=1$ in the simulations.
}
\label{MGFMfig}
\end{figure}

\section{Transverse field Ising chain}
\label{sec:Isingchain}
Let us first revisit the DQPTs in the ferromagnetic transverse field Ising chain, which will help us to understand the quench dynamics from the VBS state to the FM phase.
The Hamiltonian of transverse field Ising chain is given by
\begin{align}
H = -J_z \sum_{j} \sigma^{z}_{j} \sigma^{z}_{j+1} -h \sum_{j} \sigma^{x}_{j},
\label{Ising}
\end{align}
with the interaction strength $J_z \ge 0$ and the transverse field $h \ge 0$.
The system undergoes a second-order quantum phase transition at the critical point $h_c=1$ between the FM phase for $h < 1$ and the paramagnetic phase for $h > 1$.
In particular, the ground state is the classical ferromagnetic phase at $h=0$ and the fully polarized phase at $h \rightarrow \infty$.

We consider the quench from a fully polarized initial state, 
\begin{align}
\ket{\psi_{0}} = \bigotimes_{j=1}^{N} \frac{1}{\sqrt{2}} (\ket {\uparrow}_{j} + \ket {\downarrow}_{j}),
\label{LE:initialPolar}
\end{align}
which is the eigenstate of the Hamiltonian in Eq.(\ref{Ising}) with the transverse field $h \rightarrow \infty$, to a final Hamiltonian,
\begin{align}
H = - J_z \sum_{j} \sigma^{z}_{j} \sigma^{z}_{j+1},
\label{Isingfinal}
\end{align}
which corresponds to the Hamiltonian with the transverse field $h=0$ in Eq.(\ref{Ising}). Here $\ket{\uparrow}_{j}$, $\ket{\downarrow}_{j}$ are the two basis states of $\sigma^z_j$
denoting spin up and spin down at the $j$th site.
Then the Loschmidt amplitude $G(t)$ in Eq.(\ref{LE:LA}) can be written as,
\begin{align}
G(t) ={}& \bra{\psi_0} e^{-iHt} \ket{\psi_0} \nonumber \\
       ={}&  \bra{\psi_0} e^{iJ_{z}t \sum_{j} \sigma^{z}_{j} \sigma^{z}_{j+1}} \ket{\psi_0} \nonumber \\
       ={}& \frac{1}{2^N} \text{Tr} [e^{iJ_{z}t \sum_{j} \sigma^{z}_{j} \sigma^{z}_{j+1}}],
\label{LE:LAIsing}
\end{align}
where the $\text{Tr}$ denotes the trace. 
The Loschmidt amplitude $G(t)$ in Eq.(\ref{LE:LAIsing}) is equivalent to the partition function of the classical Ising model \cite{heyl2013dynamical,heyl2015scaling}
by replacing the time $t$ by the inverse temperature $\beta$ using $it=\beta$, then the Loschmidt amplitude $G(t)$ in Eq.(\ref{LE:LAIsing}) with PBC becomes 
\begin{align}
G(t) ={}& \text{Tr} [D^N] \nonumber \\
       ={}& \lambda_{+}^{N} + \lambda_{-}^{N}
\label{LE:LAIsingTr}
\end{align}
where $D$ is the following $2 \times 2$ matrix,
\begin{align}
D =\frac{1}{2}
\begin{pmatrix} 
e^{iJ_{z}t} & e^{-iJ_{z}t}  \\
e^{-iJ_{z}t}  & e^{iJ_{z}t}  
\end{pmatrix},
\end{align}
and $\lambda_{+}=\cos(J_{z}t)$ and $\lambda_{-}=i\sin(J_{z}t)$ are the two eigenvalues of the matrix $D$.
We can derive the critical times,
\begin{align}
 t_n=\frac{\pi}{4J_z}(2n+1),
 \label{Ising:CriticalTime}
\end{align} 
of the DQPTs by using the condition \cite{heyl2013dynamical,heyl2015scaling},
\begin{align}
|\lambda_{+}|=|\lambda_{-}|,
\end{align}
with $n$ are integers. This is equivalent to solving the equation of the Loschmidt amplitude $G(t)$,
\begin{align}
(\cos(J_{z}t))^{N} + (i\sin(J_{z}t))^{N} =0,
\end{align} 
with the condition $N=4n+2$ in the domain of real numbers. 
For $N=4n$, the critical time $t_n$ in Eq.(\ref{Ising:CriticalTime}) are obtained by finding the minima of the the Loschmidt amplitude $G(t)$, 
which decreases towards to zero when increasing system size $N$.
Hence, the rate functions in Eq.(\ref{LE:RF}) will diverge for even numbers of system sizes in the limit of $N \rightarrow \infty$, indicating DQPTs occur at critical times $t_n$.

\begin{figure}
\includegraphics[width=8.8cm]{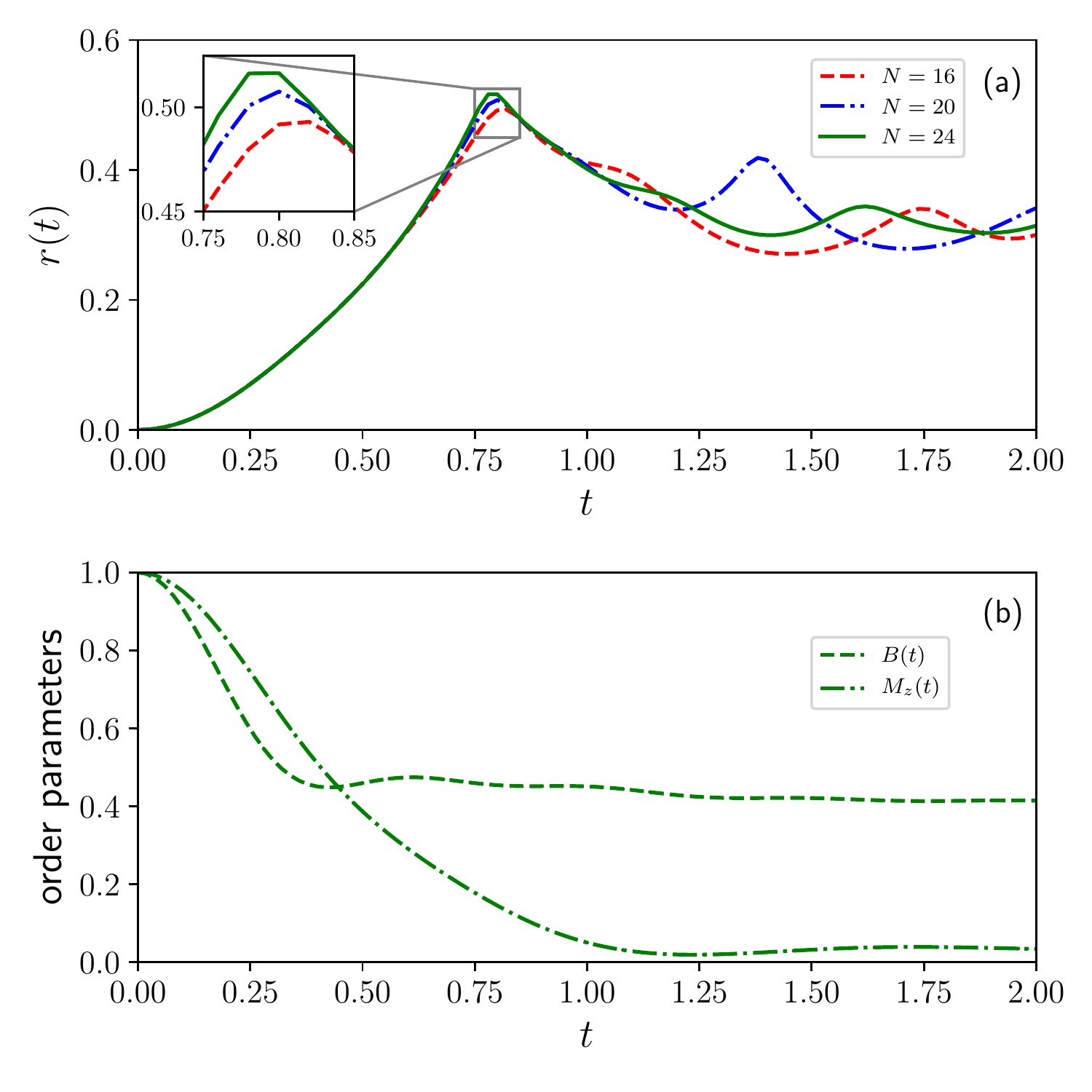}
\caption{(Color online)
Dynamics from the initial fully polarized FM phases ($N \rightarrow \infty$) to Majumdar-Ghosh phase $J_z=1$ of Eq.(\ref{Ham}) with time $t$.
(a) Rate function $r(t)$ for $N=16$ (red dashed line), $N=20$ (blue dash-dot line), $N=24$ (green solid line) lattice sites.
The inset figure in (a) denotes the finite-size effects near the first critical time.
(b) Bond order parameters $B(t)$ (dashed line) and the magnetization $M_{z}(t)$ (dash-dot line) with $N=24$ sites.
}
\label{FMtoMGfig}
\end{figure}

\section{Spin Chain with DQCP}
\label{sec:Results}
In this section, we will study the spin chain model with DQCPs defined in Eq.(\ref{Ham}) and 
analytically and numerically present our main results of DQPTs for the quenches from VBS phases to FM phases and vice versa, respectively.

For simplicity, In Eq.(\ref{Ham}) we choose $J_{x}=1$, $K_{x}=1/2$, $K_{z}=1/2$ and $J_{z}>0$ as in Ref.({\onlinecite{huang2019emergent,luo2019intrinsic,sun2019fidelity}). 
The system exhibit a phase transition from the VBS phase to the FM phase at critical point $J_z^{c} \approx 1.465$ \cite{huang2019emergent,luo2019intrinsic,sun2019fidelity}.
For $J_z < J_z^{c}$, the ground state is the VBS phase; for $J_z > J_z^{c}$, the ground state becomes the FM phase.
In particular, at $J_z=1$, the ground state is the exact Majumdar-Ghosh state \cite{Roberts2019},
\begin{align}
\ket{\psi_{0}} = \bigotimes_{m=1}^{N/2} \left(\frac{\ket {\uparrow \uparrow}+ \ket {\downarrow \downarrow}}{\sqrt{2}} \right)_{2m-1,2m},
\label{eq:InitialDM}
\end{align}
with $m$ are integers. 

We first consider the quench from the Majumdar-Ghosh state in Eq.(\ref{eq:InitialDM}) which is the eigenstate of the Hamiltonian in Eq.(\ref{Ham}) at $J_z=1$, 
to the classical Ising phase in Eq.(\ref{Isingfinal}) corresponding to the Hamiltonian in Eq.(\ref{Ham}) with $J_z \rightarrow \infty$,
The Loschmidt amplitude $G(t)$ in Eq.(\ref{LE:LA}) now becomes (see Appendix \ref{AppA} for details),
\begin{align}
G(t) ={}& \bra{\psi_0} e^{-iHt} \ket{\psi_0} \nonumber \\
       ={}& \text{Tr}[D^{N/2}] \nonumber \\
       ={}& \left(\cos(J_zt)\right)^{N/2}+\left(i\sin(J_zt)\right)^{N/2}.
\label{LE:DM}
\end{align}
This is just the Loschmidt amplitude $G(t)$ of Ising model as shown in Eq.(\ref{LE:LAIsingTr}) with the system size $N/2$.
Hence DQPTs occur at exactly the same critical times $t_n$ as in Eq.(\ref{Ising:CriticalTime}) of the transverse field Ising model.

The understanding of this result is as follows. For the case of the transverse field Ising model, the initial state in Eq.(\ref{LE:initialPolar}) is the product state of the eigenstate of single site operator $\sigma^x_j$.
In the $\sigma^z_j$ base, the eigenstate of $\sigma^x_j$ is the superposition of spin up state $\ket{\uparrow}_{j}$ and spin down state $\ket{\downarrow}_{j}$. 
Similarly, the Majumdar-Ghosh state is also a product state of entangled triplet state as shown in Eq.(\ref{eq:InitialDM}). 
And the Majumdar-Ghosh state can be regarded as a fully polarized state of system size $N/2$ if each pair in the entangled triplet state
 is grouped together as a new site $\ket{\Uparrow}_{m}=\ket{\uparrow \uparrow}_{2m-1, 2m}, \ket{\Downarrow}_{m}=\ket{\downarrow\downarrow}_{2m-1, 2m}$.

In the following, we will show such DQPTs can be described by dimerized order parameters. The dimerized order operators are defined as,
\begin{align}
D_x={}&\frac{1}{N}\sum_j(-1)^j\left(\sigma_j^x\sigma_{j+1}^x\right),\\
D_y={}&\frac{1}{N}\sum_j(-1)^j\left(\sigma_j^y\sigma_{j+1}^y\right),\\
D_z={}&\frac{1}{N}\sum_j(-1)^j\left(\sigma_j^z\sigma_{j+1}^z\right).
\end{align}
in the equilibrium VBS state. Hence, the dynamics of the dimerized order parameters are given by,
\begin{align}
D_x(t) ={}& \langle \Psi(t)|D_x|\Psi(t)\rangle, \label{LE:DMOx} \\
D_y(t) ={}& \langle \Psi(t)|D_y|\Psi(t)\rangle, \label{LE:DMOy} \\ 
D_z(t) ={}& \langle \Psi(t)|D_z|\Psi(t)\rangle, \label{LE:DMOz}
\end{align}
with $|\Psi(t)\rangle = e^{-iHt}|\Psi_0\rangle $ is the time-evolved state. 
The dimerized order parameter $D_x(t)$, $D_y(t)$, $D_z(t)$ are evaluated by (see Appendix \ref{AppB} for details),
\begin{align}
D_x(t)={}&\frac{\cos(2J_zt)^2}{2},\\
D_y(t)={}& -\frac{\cos(2J_zt)^2}{2}, \\
D_z(t)={}&\frac{1}{2}.
\end{align}
We can see that the dimerized order parameters $D_z(t)$ is conserved during the time evolution which is due to the fact that
the Majumdar-Ghosh state is the eigenstate of the quenched Hamiltonian in Eq.(\ref{Isingfinal}). 
While $D_x(t)$ and $D_y(t)$ vary sinusoidally with time $t$ in the opposite direction. 
The results of rate function $r(t)$ and the dimerized order parameters $D_x(t)$ are plotted in Fig.\ref{MGfig}, where we can clearly see the kinks in the rate functions, indicating
DQPTs occur at $t_n/t_{c1}=(2n+1)$ as predicted in Eq.(\ref{Ising:CriticalTime}), with the $t_{c1}=\frac{\pi}{4}$ being the first critical time. 
We find that DQPTs can be described by the x-component dimerized order parameters $D_x(t)$ (or y-component $D_y(t)$) which becomes zero at critical time $t_n$ \cite{hagymasi2019dynamical}. 
Interestingly, the value of the x-component dimerized order parameters $D_x(t)$ is just the half of the value of $\langle\sigma^x(t)\rangle$ in transverse field Ising chain. 
It supported our argument again that we can group the two entangled particles together and consider its dynamics as that in Ising model. 

Indeed, above arguments and results of the Loschmidt amplitude $G(t)$ can be generated to any product state consisting of $K$ qubits Greenberger-Horne-Zeilinger (GHZ) state,
\begin{align}
\ket{\psi_{0}} = \bigotimes_{m=1}^{N/K} \left(\frac{\ket {\uparrow}^{\otimes K}+ \ket {\downarrow}^{\otimes K}}{\sqrt{2}} \right)_{Km-K+1,\dotsb,Km}.
\label{eq:CatState}
\end{align}
The Loschmidt amplitude $G(t)$ for initial $K$ qubits GHZ state would become,
\begin{align}
G(t)={}& \left(\cos(J_zt)\right)^{N/K}+\left(i\sin(J_zt)\right)^{N/K},
\label{LE:GHZ}
\end{align}
if we group the $K$ qubits GHZ state as a new single lattice size.
We numerically confirm above analytical results of Loschmidt amplitude $G(t)$ by using the exact diagonalization for GHZ states in small systems.

Next let us start to consider the general quenches from the Majumdar-Ghosh state $J_z=1$ to FM states (large but finite $J_z$) by decreasing the $J_z$ from $J_z \rightarrow \infty$.
For the very large $J_z$, where the fluctuations are weak, we expect that the DQPTs will survive. To support our argument, we perform the exact diagonalization
up to $N=24$ lattice size for $J_z/J_x=50$, $J_z/J_x=20$ and $J_z/J_x=10$ in PBCs, respectively. The rate function $r(t)$ and x-component
dimerized order parameters $D_x(t)$ are presented in Fig.\ref{MGFMfig}. For very large $J_z=50$, the rate function $r(t)$ and order parameters $D_x(t)$ 
change little compared to the classical Ising model. Increasing $J_z$, the peak of rate functions $r(t)$ and minima of order parameters $D_x(t)$ move to the right direction
due to the stronger fluctuations in quenched Hamiltonian. 
We find the peaks of the rate functions $r(t)$ increase with the system sizes around the local minima of the order parameters $D_x(t)$ indicating that DQPTs persist even for $J_z/J_x=10$ (see Fig.\ref{MGFMfig}c). 
And the existence of DQPTs are robust under a small perturbation to the initial Majumdar-Ghosh state (i.e. changing $J_z/J_x = 1$ to $J_z/J_x = 1.05$, see Fig.\ref{MGFMfig}a).
We note that when $J_z$ is close to the equilibrium critical point $J_z^c \approx 1.465$, we cannot find nice kinks due to the strong fluctuations where the criticality
of the DQCP will play an important role in the DQPTs. The study on the quenches near the critical point is a very difficult problem that we leave for the future work.

Finally, we will briefly discuss the DQPTs from the quenches from fully polarized FM phase ($J_z \rightarrow \infty$) to the Majumdar-Ghosh state ($J_z=1$).
We quench our system from one of the following doubly degenerate polarized FM phase,
\begin{align}
\ket{\psi_{0}} = \bigotimes_{i=1}^{N}  (\ket {\uparrow}_{i} ).
\end{align}
to the Majumdar-Ghosh model and perform the exact diagonalization to compute the rate functions $r(t)$, bond order parameters $B(t)=\langle \vec{\sigma}_{i} \cdot \vec{\sigma}_{i+1}\rangle$,
and the magnetization $M_{z}(t)=\frac{1}{N} \sum_{i} \langle \sigma_i^z\rangle$ in PBCs. 
The results are shown in Fig.\ref{FMtoMGfig}, where we can see that the bond order parameters $B(t)$ and the magnetization $M_{z}(t)$
decay very quickly to equilibrium values of $B(t) \approx 1/2$ and $M_{z}(t) \approx 0$ so that it is very difficult to denote the DQPTs although it seem that there is a DQPT (kink structure) in the short time scale.
We note that our result in this case is different from that in the XXZ model \cite{heyl2014dynamical}, where the magnetization shows an oscillatory behavior 
and DQPTs can be well described by comparing two rate functions $r_{\eta}(t)$, with $\eta$ denoting two degenerate N{\'e}el phase.
Therefore, our results reveal that different broken symmetries will play a different role in the quench dynamics. It would very interesting to understand the relations between DQPTs and symmetries in the future.

\section {Conclusion}
\label{sec:Con}
In this paper, we study the quench dynamics in a spin chain with a DQCP. We drive analytical results of Loschmidt amplitude and order parameters
for the quench from Majumdar-Ghosh state to the classical Ising chain. For more general cases, we numerically investigate the quench dynamics. 
We show that DQPTs can occur in systems with two broken symmetry phases and can be described by x-component (or y-component) dimerized VBS order parameters.
Our results reveal that broken lattice symmetry and broken spin symmetry of quenched Hamiltionian play a different role in the quench dynamics. 
For the quench from the broken lattice symmetry to the $\mathbb{Z}_2$ broken classical Ising chain, we find that the dynamics of Loschmidt amplitude with initial Majumdar-Ghosh state  
is equivalent to a product state of a translation symmetry. This means we cannot distinguish the Ising transition and the DQCP from such quench dynamics, implying that 
one should consider the quenches near the DQCP in order to study its critical properties \cite{zhou2019signature,hwang2019universality,ding2020dynamical}. 
We note that our results for the initial VBS states and any $K$ quibits GHZ states may be realized in recent experiments \cite{song2019generation}.

It would be very interesting to investigate the quench dynamics in two-dimensional systems with DQCPs to know whether DQPTs can occur and whether the dynamics of Loschmidt amplitude
can be mapped to two dimensional classical Ising model.

\begin{acknowledgments}
We would like to thank M. Heyl for useful correspondence on DQPTs.
G. S. is appreciative of support from the NSFC under the Grant No. 11704186 and the startup Fund of Nanjing University of Aeronautics and Astronautics under the Grant No. YAH17053.
B. B. W. is appreciative of support from the NSFC under the Grant No. 11604220 and the President's Fund of The Chinese University of Hong Kong, Shenzhen.
Numerical simulations were carried out on the clusters at Nanjing University of Aeronautics and Astronautics.
\end{acknowledgments}

\appendix
\section{Derivation of Loschmidt amplitude}
\label{AppA}
To investigate the DQPTs in the system, we start from the dimerized Majumdar-Ghosh state which is given by
\begin{eqnarray}
|\Psi_0\rangle=\bigotimes_{m=1}^{N/2} \left(\frac{|\uparrow\uparrow\rangle+|\downarrow\downarrow\rangle}{\sqrt{2}}\right)_{2m-1,2m},
\label{AppAMG}
\end{eqnarray}
where $|\uparrow,\downarrow\rangle$ are the basis state along $z$ direction. We then quench this state by a Hamiltonian deep in the z-FM regime with $J_z\gg1$, which is
\begin{eqnarray}
H=-\sum_{i=1}^N J_{z} \sigma^{z}_{i} \sigma^{z}_{i+1}.
\end{eqnarray}
To evaluate the Loschmidt echo, we note that,
\begin{eqnarray}
|\Psi(t)\rangle &=& e^{-itH}|\Psi_0\rangle, \nonumber \\
&=&\prod_ie^{itJ_z\sigma_i^z\sigma_{i+1}^z} |\Psi_0\rangle, \nonumber \\
&=&e^{itJ_zN/2}\prod_{i=2,4,\cdots,2m}e^{itJ_z\sigma_i^z\sigma_{i+1}^z} |\Psi_0\rangle. 
\end{eqnarray}
If we group each pair in the entangled triplet state as a new site, 
 \begin{eqnarray}
 \ket{\Uparrow}_{m}=\ket{\uparrow \uparrow}_{2m-1, 2m}, \ket{\Downarrow}_{m}=\ket{\downarrow\downarrow}_{2m-1, 2m}
 \end{eqnarray}
the Majumdar-Ghosh state in Eq.(\ref{AppAMG}) becomes,
\begin{eqnarray}
\ket{\psi_{0}} = \bigotimes_{m=1}^{N/2} \frac{1}{\sqrt{2}} (\ket {\Uparrow}_{m} + \ket {\Downarrow}_{m}),
\label{LE:AppAPolar}
\end{eqnarray}
The eigenvalue equation of operators,
\begin{eqnarray}
\sigma_{2m}^z = \mathbb{1}_{2m-1} \otimes \sigma_{2m}^z, \\
\sigma_{2m+1}^z =\sigma_{2m+1}^z \otimes \mathbb{1}_{2m+2},
\end{eqnarray}
are,
\begin{eqnarray}
\sigma_{2m}^z  \ket{\psi_{0}} = \frac{1}{\sqrt{2}} (\ket {\Uparrow}_{m} - \ket {\Downarrow}_{m}), \\
\sigma_{2m+1}^z  \ket{\psi_{0}} = \frac{1}{\sqrt{2}} (\ket {\Uparrow}_{m} - \ket {\Downarrow}_{m}),
\end{eqnarray}
which are the same as the Ising model. Then the Loschmidt amplitude for PBC is,
\begin{eqnarray}
G(t)&=&\langle\Psi_0|e^{-itH}|\Psi_0\rangle, \nonumber \\
&=&\langle\Psi_0| e^{itJ_zN/2}\prod_{i=2,4,\cdots,2m}e^{itJ_z\sigma_i^z\sigma_{i+1}^z} |\Psi_0\rangle, \nonumber \\
&=&\text{Tr}[D^{N/2}], \nonumber \\
&=&\left(\cos(J_zt)\right)^{N/2}+\left(i\sin(J_zt)\right)^{N/2}.
\end{eqnarray}

\section{Derivation of order parameters}
\label{AppB}
Now let us first see how the polarizations $\langle\sigma^x\rangle$ in $x$ direction evolves in classical Ising model for PBC,
\begin{eqnarray}
\langle\sigma^x(t)\rangle&=&\frac{1}{N}\langle\sum_j\sigma_j^x\rangle, \nonumber \\
&=&\frac{1}{N}\langle\Psi_0|e^{-iJ_zt\sum_i\sigma_i^z\sigma_{i+1}^{z}}\sum_j\sigma_j^xe^{iJ_zt\sum_i\sigma_i^z\sigma_{i+1}^{z}}|\Psi_0\rangle, \nonumber \\
&=&\langle\Psi_0|e^{-iJ_zt\sum_i\sigma_i^z\sigma_{i+1}^{z}}\sigma_1^xe^{iJ_zt\sum_i\sigma_i^z\sigma_{i+1}^{z}}|\Psi_0\rangle, \nonumber \\
&=&\langle\Psi_0|e^{-iJ_zt\sum_{i=2}^{N-1}\sigma_i^z\sigma_{i+1}^{z}}e^{-iJ_zt\sigma_1^z(\sigma_N^z+\sigma_2^z)}\sigma_1^x \nonumber \\
&& e^{iJ_zt\sigma_1^z(\sigma_N^z+\sigma_2^z)}e^{iJ_zt\sum_{i=2}^{N-1}\sigma_i^z\sigma_{i+1}^{z}}|\Psi_0\rangle, \nonumber \\
&=&\langle\Psi_0|e^{-iJ_zt\sum_{i=2}^{N-1}\sigma_i^z\sigma_{i+1}^{z}}(\sigma_1^x\cos[2J_zt(\sigma_N^z+\sigma_2^z)] \nonumber \\
&& +\sigma_1^y\sin[2J_zt(\sigma_N^z+\sigma_2^z)])e^{iJ_zt\sum_{i=2}^{N-1}\sigma_i^z\sigma_{i+1}^{z}}|\Psi_0\rangle, \nonumber \\
&=&\langle\Psi_0|e^{-iJ_zt\sum_{i=2}^{N-1}\sigma_i^z\sigma_{i+1}^{z}}(\cos[2J_zt(\sigma_N^z+\sigma_2^z)]) \nonumber \\
&& e^{iJ_zt\sum_{i=2}^{N-1}\sigma_i^z\sigma_{i+1}^{z}}|\Psi_0\rangle, \nonumber \\
&=&\cos^2(2J_zt).
\end{eqnarray}

Next, we will show the details computing the dimerized order parameters $\langle D_x(t)\rangle$, $\langle D_y(t)\rangle$, $\langle D_z(t)\rangle$ 
defined in Eq.(\ref{LE:DMOx}), Eq.(\ref{LE:DMOy}) and Eq.(\ref{LE:DMOz}).
The order parameter dynamics $\langle D_z(t)\rangle$ can be easily calculated as
\begin{eqnarray}
\langle D_z(t)\rangle&=&\langle\Psi_0|e^{itH}D_ze^{-itH}|\Psi_0\rangle, \nonumber \\
&=&\langle\Psi_0|D_z|\Psi_0\rangle, \nonumber \\
&=&\frac{1}{2}.
\end{eqnarray}
To evaluate $\langle D_x(t)\rangle$ and $\langle D_y(t)\rangle$, we note that the quenched state is of the form,
\begin{eqnarray}
|\Psi(t)\rangle &=& e^{itJ_zN/2} e^{ i J_z t \sum\limits_{i=2,4,\cdots,2m} \sigma_i^z\sigma_{i+1}^z} |\Psi_0\rangle. 
\end{eqnarray}
Thus we have,
\begin{eqnarray}
&& \langle\Psi(t)|\sigma_1^x\sigma_2^x|\Psi(t)\rangle \nonumber \\
&=& \langle\Psi_0|e^{-iJ_zt\sum\limits_{i=\text{even}}\sigma_i^z\sigma_{i+1}^{z}}\sigma_1^x\sigma_2^xe^{iJ_zt\sum\limits_{i=\text{even}}\sigma_i^z\sigma_{i+1}^{z}}|\Psi_0\rangle, \nonumber \\
&=& \langle\Psi_0|e^{-iJ_zt\sum_{i=4}^{2m-2}\sigma_i^z\sigma_{i+1}^{z}}e^{-iJ_zt(\sigma_2^z\sigma_3^z+\sigma_1^z\sigma_N^z)}\sigma_1^x\sigma_2^x \nonumber \\
&& e^{iJ_zt(\sigma_1^z\sigma_N^z+\sigma_2^z\sigma_3^z)}e^{iJ_zt\sum_{i=4}^{2m-2}\sigma_i^z\sigma_{i+1}^{z}}|\Psi_0\rangle, \nonumber \\
&=& \langle\Psi_0|e^{-iJ_zt\sum_{i=4}^{2m-2}\sigma_i^z\sigma_i^{z+1}}(\sigma_1^x\sigma_2^x\cos[2J_zt(\sigma_N^z)\cos[2J_zt(\sigma_3^z)] \nonumber \\
&& +\sigma_1^y\sigma_2^y\sin[2J_zt(\sigma_N^z)\sin[2J_zt(\sigma_3^z)])e^{iJ_zt\sum_{i=4}^{2m-2}\sigma_i^z\sigma_i^{z+1}}|\Psi_0\rangle, \nonumber \\
&=& \cos^2(2J_zt).
\end{eqnarray}
Similarly we have
\begin{eqnarray}
\langle\Psi(t)|\sigma_2^x\sigma_3^x|\Psi(t)\rangle&=&0,\\
\langle\Psi(t)|\sigma_3^x\sigma_4^x|\Psi(t)\rangle&=&\cos^2(2J_zt),\\
\langle\Psi(t)|\sigma_4^x\sigma_5^x|\Psi(t)\rangle&=&0.
\end{eqnarray}
Finally we get
\begin{eqnarray}
\langle D_x(t)\rangle&=& \langle \Psi(t)|D_x|\Psi(t)\rangle \nonumber \\
&=&\frac{\cos^2(2J_zt)}{2},\\
\langle D_y(t)\rangle&=& \langle \Psi(t)|D_y|\Psi(t)\rangle \nonumber \\ 
&=& -\frac{\cos^2(2J_zt)}{2}.
\end{eqnarray}

\bibliographystyle{apsrev4-1}
\bibliography{ref}

\end{document}